\documentclass[12pt,letterpaper]{article}
\usepackage[margin=1in]{geometry}
\usepackage{setspace}
\usepackage{amsmath,amssymb,amsthm}
\usepackage[authoryear,round]{natbib}
\usepackage{graphicx,psfrag,blkarray}
\usepackage{caption,subcaption,multirow,enumerate}
\usepackage{bm,xcolor,tikz}
\usepackage{booktabs,makecell,longtable,array,tabularx,threeparttable}
\usepackage{url}
\usepackage[ruled,noend]{algorithm2e}
\usepackage[hidelinks]{hyperref}

\theoremstyle{plain}
\newtheorem{theorem}{Theorem}
\newtheorem{lemma}{Lemma}

\theoremstyle{definition}
\newtheorem{definition}{Definition}
\newtheorem{assumption}{Assumption}

\theoremstyle{remark}

\newcommand{\bpi}{\bm{\pi}}

\definecolor{darkgreen}{RGB}{0,100,0}

\graphicspath{{./art/}{./Figure/}}

\newcommand{\blind}{1}
\title{\bfseries From multilayer networks to covariates: Layer-specific community detection}
\author{Tao Shen and Wanjie Wang\thanks{%
Tao Shen is at the Department of Statistics and Data Science,
National University of Singapore, 6 Science Drive 2,
Singapore 117546, Singapore (email: \texttt{taoshen@u.nus.edu}).
Wanjie Wang is at the Department of Statistics and Data Science
\& Department of Mathematics, National University of Singapore,
6 Science Drive 2, Singapore 117546, Singapore
(email: \texttt{wanjie.wang@nus.edu.sg}).}}
\date{}

\begin{document}

\begin{singlespace}
\if1\blind
  \maketitle
\else
  \begin{center}
    {\LARGE\bfseries From multilayer networks to covariates: Layer-specific community detection\par}
  \end{center}
  \medskip
\fi
\end{singlespace}

\begin{abstract}
We study layer-specific community detection in multilayer networks with heterogeneous community memberships and degree distributions. 
Membership mismatch complicates cross-layer information pooling, whereas sparsity heterogeneity can provide complementary signals for poorly connected communities. 
We recast the problem as network-with-covariates clustering: the target layer is retained as the primary network, while the remaining layers are converted into nodal covariates through our regularized spectral construction approach. 
We then perform community detection using a node-adaptive balancing scheme that relies more on the target network for high-degree nodes and more on the constructed covariates for low-degree nodes. 
Under a multilayer degree-corrected stochastic blockmodel, we establish guarantees for the constructed covariates and weak consistency when the average membership mismatch vanishes. Under additional degree structure and node-specific mismatch conditions, we establish strong consistency while allowing the average membership mismatch to remain nonvanishing. 
Simulations show that our proposed approach achieves the lowest misclustering error rates under most scenarios. An application to international food-trading networks reveals interpretable product-specific community structures.
\end{abstract}

\par\bigskip

\section{Introduction}
\label{sec-intro}

Multilayer networks represent multiple relational sources on a common set of nodes, with different layers corresponding to distinct interactions or networks constructed from different data sources \citep{kivela2014multilayer, banerjee2013diffusion, zhu2022dea}. 
An $L$-layer network on $n$ nodes can be denoted by adjacency matrices $\{A^{(l)} \in \{0, 1\}^{n \times n}\}_{l\in[L]}$.
These layers need not encode exactly the same structural information.
For example, let $\pi^{(l)}\in[K]^n$ denote the unknown community memberships in layer $l$. Then $\pi^{(l)}(i)$ for the same node $i$ may be different across layers $l \in [L]$ \citep{wang2019common, zhang2020flexible}. Such layer-to-layer variation may itself be scientifically meaningful. Our target is therefore to recover layer-specific memberships $\pi^{(l)}$, $l \in [L]$, rather than a common partition shared by all layers.

The heterogeneity is present not only in the layer-to-layer memberships, but also in node degrees, which can vary substantially across nodes and layers. 
Some nodes may have extremely low degrees in a given target layer, making their community labels difficult to infer from that layer alone. At the same time, cross-layer degree heterogeneity means that the same nodes or the corresponding communities may be much better connected in other layers.
Community detection based on a single target layer preserves its layer-specific structure but may fail for extremely low-degree nodes \citep{qin2013regularized, lei2015consistency, jin2015fast}. Conversely, pooling layers can borrow information from better-connected layers \citep{han2015consistent, paul2020spectral, lei2020consistent}, but may obscure the target-layer memberships under cross-layer membership heterogeneity.

The membership and degree heterogeneity play distinct roles in this problem. Membership heterogeneity creates cross-layer mismatch, so using auxiliary layers can introduce error \citep{chen2022global, lei2023bias, agterberg2025joint}. 
In contrast, degree heterogeneity creates potential complementarity: nodes with very low degrees in the target layer may be much better connected in other layers.
Thus, heterogeneity raises new challenges but also brings new opportunities. For the target layer, our goal is to use this cross-layer complementarity to improve community recovery for sparse target-layer nodes, while controlling the error caused by membership mismatch. 

Our key idea is to adopt a primary-auxiliary view of the multilayer network.
For a target layer $l$, we retain its adjacency matrix $A^{(l)}$ as the primary network and convert the auxiliary layers into a matrix of nodal covariates $X^{(l)}$. 
This recasts the original multilayer problem as a network-with-covariates problem on $(A^{(l)}, X^{(l)})$. The two components play distinct roles: $A^{(l)}$ specifies the community structure to be recovered, while $X^{(l)}$ carries complementary information from the auxiliary layers. This representation provides a framework in which cross-layer membership mismatch can be accommodated while complementarity arising from degree heterogeneity can be exploited for layer-specific community recovery.

This primary--auxiliary framework relies on how $X^{(l)}$ is constructed. In particular, the construction must accommodate severe degree heterogeneity across nodes, communities, and layers.
We propose \emph{multilayer-assisted regularized covariates} (MARC) for this purpose. For each auxiliary layer, MARC forms an eigenvalue-weighted and degree-regularized spectral representation. Eigenvalue weighting downweights noisy weak spectral directions, while degree regularization mitigates node-level degree variation and instability from extremely sparse nodes. The resulting representations are stacked to form $X^{(l)}$, combining complementary community information across the auxiliary layers. 

We then develop a layer-specific community detection approach based on $(A^{(l)}, X^{(l)})$. 
Inspired by the balancing design in \cite{hu2024network}, we introduce a node-specific diagonal weight matrix $\beta^{(l)}$ and construct $Y^{(l)} = A^{(l)} X^{(l)} + \beta^{(l)} X^{(l)}$. 
The weights adapt to the target-layer degrees, allowing the constructed covariates to play a dominant role for extremely low-degree nodes and the target network for nodes with higher degrees. 
Spectral clustering on $Y^{(l)}$ then estimates the community memberships of layer $l$. 

We establish both weak and strong consistency for the layer-specific community estimates under a multilayer degree-corrected stochastic blockmodel. We characterize how average membership mismatch and degree heterogeneity affect community recovery, and establish weak consistency when the former vanishes. 
Under additional degree structure and node-specific mismatch conditions, we establish strong consistency while allowing a sufficiently small but nonvanishing average mismatch. 

Our setting lies between two common strategies for multilayer community detection. 
One is to analyze each layer separately using single-network methods, including spectral methods for stochastic block models and their degree-corrected extensions \citep{qin2013regularized, lei2015consistency, jin2015fast, jing2022community}. This allows community memberships to vary across layers, but requires each layer to be sufficiently connected for reliable recovery. 
The other strategy pools information across layers, typically under a common community structure. Examples include likelihood-based estimation \citep{han2015consistent}, adjacency aggregation \citep{bhattacharyya2017spectral, lei2023bias}, and aggregation of spectral information \citep{paul2020spectral, arroyo2021inference, agterberg2025joint}. 
By combining multiple layers, these methods can relax the connectivity requirement on individual layers, but their common-structure assumption can be restrictive when community memberships vary across layers. 


Several methods relax the common-membership assumption by allowing community structures to vary across layers. \citet{jing2021community} group layers into network types with type-specific community memberships and aggregate layers within each type. \citet{chen2022global} consider a two-community model, in which each layer-specific membership is a perturbation of a global membership, and refine the global and layer-specific memberships after spectral clustering. Latent-space approaches instead represent cross-layer heterogeneity through common and layer-specific latent coordinates \citep{macdonald2022latent, tian2024efficient}. These methods allow community structures to differ across layers, but typically require nontrivial degree conditions for individual layers or cross-layer alignment conditions. They also do not directly address how to exploit degree heterogeneity when some nodes or communities are extremely sparse in one layer but better connected in others. 

A complementary perspective comes from community detection with auxiliary nodal covariates \citep{newman2016structure, binkiewicz2017covariate, yan2021covariate, hu2024network}, with theoretical guarantees for spectral methods \citep{abbe2022l, hu2024network}. These methods typically treat the covariates as externally observed inputs, rather than constructing them from other network layers. In our setting, the construction itself is nontrivial because their memberships and degree distributions may differ from those of the target layer. Once constructed, these covariates must also be combined with the target network in a way that accounts for such heterogeneity. We build on the balancing design in \cite{hu2024network} and combine it with MARC covariates for layer-specific community detection.

Against this background, our contributions are summarized as follows.
\begin{itemize}
\item First, we introduce a primary--auxiliary formulation for layer-specific inference in heterogeneous multilayer networks and propose MARC to implement it. To our knowledge, this is the first approach that converts heterogeneous auxiliary network layers into target-specific nodal covariates while accommodating both cross-layer membership and degree heterogeneity. For each target layer, MARC constructs the covariates through eigenvalue-weighted and degree-regularized spectral representations. 

\item Second, we develop a layer-specific community detection procedure based on the target network and the MARC covariates. A node-adaptive weighting scheme balances the two components according to target layer degrees, allowing the covariates to play a dominant role for extremely low-degree nodes and the target network for nodes with higher degrees.

\item Third, we provide the first theoretical analysis of multilayer community detection that simultaneously accommodates cross-layer membership mismatch and extreme sparsity. Our framework allows a constant proportion of nodes to have bounded population degrees in individual layers.
We derive an explicit error bound for MARC. Building on this bound, we establish weak consistency under vanishing average mismatch and strong consistency under additional degree-structure and node-specific conditions, allowing sufficiently small but nonvanishing average mismatch.

\item Finally, we provide an optional weighted consensus of the layer-specific estimates. This consensus consistently recovers an approximately common reference partition under suitable recovery and identification conditions.
\end{itemize}

Simulation studies in Section \ref{sec-simu} show that MARC improves both the proposed community-detection procedure and existing network-with-covariates methods, particularly when the target layers are sparse and the community memberships vary across layers. The international food-trading networks analysis in Section \ref{sec-real} further illustrates how the resulting layer-specific communities reveal product-dependent trading structures.


\section{Multilayer-assisted covariate construction and community detection}
\label{sec-method}

\subsection{Heterogeneous multilayer networks}
\label{sec-pre}
Consider an $L$-layer network $\{A^{(l)}\}_{l\in[L]}$ on a common set of $n$ nodes, where $A^{(l)}\in\{0,1\}^{n\times n}$ is the symmetric adjacency matrix for layer $l$, with $A^{(l)}_{ij}=1$ indicating an edge between nodes $i$ and $j$. 
Suppose that the nodes form $K$ communities in each layer, where $K$ is known.
Let $\pi^{(l)}\in[K]^n$ denote the community membership vector for layer $l$, with
$\pi^{(l)}(i)$ giving the membership of node $i$.
Our target is to recover the collection of layer-specific community memberships
$\{\pi^{(l)}\}_{l\in[L]}$.

We first quantify the heterogeneity in cross-layer community assignment. Let $\pi^{(l)}$ be the membership vector for layer $l$, and we define the pairwise mismatch matrix
\[
Q=(q_{ls})_{l,s\in[L]},
\qquad
q_{ls}
=
\frac{1}{n}
\min_{\kappa\in\mathfrak S_K}
\sum_{i=1}^n
I\left\{
\pi^{(l)}(i)
\neq
\kappa\bigl(\pi^{(s)}(i)\bigr)
\right\},
\]
where $\mathfrak S_K$ denotes the set of permutations of $[K]$. Thus, $q_{ls}$ measures the proportion of nodes with different community assignments in layers $l$ and $s$, up to a permutation.

To model degree heterogeneity, we adopt the multilayer degree-corrected stochastic block model (multi-DCSBM) as a working model \citep{karrer2011stochastic,zhao2012consistency,yan2014model}.
For layer $l$, let $\Pi^{(l)}\in\{0,1\}^{n\times K}$ denote the membership matrix, where $\Pi^{(l)}_{ik}=1$ if and only if $\pi^{(l)}(i)=k$, and let $\Theta^{(l)}=\operatorname{diag}(\theta^{(l)}_1,\ldots,\theta^{(l)}_n)$, where $\theta^{(l)}_i$ is the degree heterogeneity parameter of node $i$.
Let $B^{(l)}\in\mathbb R^{K\times K}$ denote the block connectivity matrix.
Conditional on $\Pi^{(l)}$, the upper-triangular entries of $A^{(l)}$ are independent Bernoulli variables with
\begin{align}
E[A^{(l)}\mid\Pi^{(l)}]
&=
A^{(l)*}-\operatorname{diag}(A^{(l)*}),
\qquad
A^{(l)*}
=
\Theta^{(l)}\Pi^{(l)}B^{(l)}
(\Pi^{(l)})^\top\Theta^{(l)} .
\label{eq:mdcsbm}
\end{align}
Existing multilayer block-model formulations commonly impose a shared community assignment $\pi$ across layers, while allowing $B^{(l)}$ or $\Theta^{(l)}$ to vary \citep{paul2020spectral,lei2023bias,agterberg2025joint,zhao2025spectral}.
Our work allows $\Pi^{(l)}$, $\Theta^{(l)}$, and $B^{(l)}$ to vary with $l$ simultaneously, so that both community memberships and degree profiles can be
layer-specific.


Let
$d_i^{(l)}=\sum_{j\in[n]}A^{(l)}_{ij}$ and
$d_i^{(l)*}=\sum_{j\neq i}A_{ij}^{(l)*}$
denote the observed and population degrees of node $i$ in layer $l$, respectively.
We focus on two sparsity regimes, while allowing an intermediate range between them. Nodes with population degree at least of logarithmic order, $d_i^{(l)*}\gtrsim \log n$, are referred to as relatively sparse. 
This logarithmic degree scale is considered the usual sparse-network regime in community detection theory to guarantee strong consistency \citep{qin2013regularized, hu2024network}; the term ``relatively sparse'' is used here only in contrast to the more severe regime below and may also include nodes with larger degrees. In comparison, we refer to nodes satisfying
\(
d_i^{(l)*}
=
o\!\left(\frac{\log n}{\log\log n}\right)
\)
as {\it extremely sparse}. Their population degrees grow much more slowly than the usual logarithmic scale and may be of constant order or even zero. The information from extremely sparse nodes is insufficient for label recovery; see \cite{hu2024network}. Nodes between these two scales form an intermediate regime, for which our weighting rule provides a smooth transition between the two cases.
The formal community-level definitions of these sparsity regimes are given in Section~\ref{sec-theory}.
Putting these together, our setting allows both community memberships and degree heterogeneity to vary across layers, where the former quantified by $Q$ and the latter allowing extremely sparse nodes. Their combination makes layer-specific community detection particularly challenging. We would like to leverage all layers, while carefully controlling the error caused by community misalignment.



\subsection{MARC: Multilayer-assisted regularized covariates}
\label{sec-mainalgo}
We recast the multilayer network into a collection of network-with-covariates pairs $(A^{(l)}, X^{(l)})$, $l \in [L]$. 
For a target layer $l$, we retain $A^{(l)}$ as the primary network and convert the remaining layers $\{A^{(s)}\}_{s \neq l}$ into a nodal covariate matrix $X^{(l)}$. 
This conversion preserves the target layer as the primary inferential object, while allowing the auxiliary layers to contribute additional community information. 
It translates cross-layer membership heterogeneity into a network-covariate mismatch, which is addressed by the construction of $X^{(l)}$ and the subsequent network-with-covariates clustering procedure.

We propose a novel covariate construction, MARC (\emph{multilayer-assisted regularized covariates}), to construct $X^{(l)}$.
MARC is motivated by the population spectral structure of the auxiliary layers under the multi-DCSBM. For an auxiliary layer $s\neq l$, let
$A^{(s)*}
=
U^{(s)*}\Sigma^{(s)*}(U^{(s)*})^\top$,
and let $U_K^{(s)*}\Sigma_K^{(s)*}$ denote the weighted population embedding corresponding to the $K$ eigenvalues with largest magnitudes.
Under the multi-DCSBM, there exists a community-level matrix
$V^{(s)}\in\mathbb R^{K\times K}$ such that
\begin{equation}
e_i^\top U_K^{(s)*}\Sigma_K^{(s)*}
=
\theta_i^{(s)}
e_{\pi^{(s)}(i)}^\top V^{(s)}.
\label{eq:population-weighted-embedding}
\end{equation}
Hence, at the population level, nodes in the same community share a common spectral direction, up to a row-wise degree heterogeneity parameter \citep{qin2013regularized,jin2015fast, lei2015consistency}. 

Motivated by this population structure, we construct an empirical
covariate block from each auxiliary layer. For $s\neq l$, let $A^{(s)} = U^{(s)}\Sigma^{(s)}(U^{(s)})^\top$ be the eigendecomposition of $A^{(s)}$, and let $U_{\widehat K}^{(s)}\Sigma_{\widehat K}^{(s)} \in\mathbb R^{n\times\widehat K}$
denote the weighted spectral embedding corresponding to the $\widehat K$ eigenvalues with largest magnitudes. 
Here, $\widehat K$ is the number of spectral directions retained in the construction of $X^{(l)}$, which may differ from the number of communities $K$.
We then deal with the degree heterogeneity in $U_{\widehat K}^{(s)}\Sigma_{\widehat K}^{(s)}$. 
Let the degree diagonal matrix 
$D^{(s)} = \operatorname{diag}(d_1^{(s)},\ldots,d_n^{(s)})$.
We set the regularization level to $\tau_n = c\log n$, and use $c=5$ in all simulations and the real-data analysis. Define the regularized covariate block from layer $s$,
\[
S^{(s)}
=
\{D^{(s)}+\tau_n\, I_n\}^{-1}
U_{\widehat K}^{(s)}\Sigma_{\widehat K}^{(s)}.
\]
For target layer $l$, we stack the auxiliary-layer blocks to construct
the covariate matrix
\[
X^{(l)}
=
[S^{(1)},\ldots,S^{(l-1)},S^{(l+1)},\ldots,S^{(L)}]
\in\mathbb R^{n\times (L-1)\widehat K}.
\]

Our MARC construction combines three operations. First, we weight the spectral directions by $\Sigma_{\widehat K}^{(s)}$ to accommodate cross-layer degree heterogeneity. Denser communities tend to induce spectral directions with larger eigenvalue magnitudes, which are typically more reliably estimated, whereas sparser communities tend to contribute directions with smaller eigenvalue magnitudes and greater estimation uncertainty. Eigenvalue weighting therefore places more weight on the more reliable spectral directions and less weight on the less reliable ones, which deals with cross-layer degree heterogeneity. 
Second, we apply the degree-regularized row scaling $\{D^{(s)}+\tau_n\,I_n\}^{-1}$ to accommodate node-specific degree heterogeneity within each layer. 
As shown in \eqref{eq:population-weighted-embedding}, the population spectral embedding depends on the degree parameter $\theta_i^{(s)}$.
For relatively sparse nodes, $d_i^{(s)}+\tau_n $ is of the same order as $ d_i^{(s)}$, providing a stable scale for this $\theta_i^{(s)}$ effect, so scaling by $(d_i^{(s)}+\tau_n)^{-1}$ approximately removes the variation induced by $\theta_i^{(s)}$. 
For extremely sparse nodes, the observed degree can be too small and unstable for direct normalization. The regularization term $\tau_n$ then prevents excessive rescaling and keeps the corresponding row magnitudes controlled.
Finally, we stack the blocks $\{S^{(s)}\}_{s\neq l}$ so that each auxiliary layer enters through its own block, without requiring the estimation or alignment of community memberships across layers \citep{arroyo2021inference,agterberg2025joint}.

The construction does not require a one-to-one correspondence between the retained empirical spectral directions and the community labels. Accordingly, the retained spectral dimension $\widehat K$ need not coincide with the number of communities $K$.
In practice, moderate over-specification with $\widehat K\ge K$ is preferable to under-specification, since the eigenvalue weighting
attenuates weaker retained directions, whereas $\widehat K<K$ may discard informative ones \citep{shen2025optimal}.
The MARC construction is summarized in Algorithm~\ref{algo-cc}.

\begin{algorithm}[htbp]
\caption{Multilayer-assisted regularized covariates (MARC)}
\label{algo-cc}
\KwIn{Adjacency matrices
$\{A^{(s)}\}_{s\in[L]\setminus\{l\}}$ and spectral dimension
$\widehat K$}

\For{$s\in[L]\setminus\{l\}$}{
  Compute the $\widehat{K}$ eigenpairs with largest eigenvalue magnitudes $U^{(s)}_{\widehat{K}}$ and $\Sigma^{(s)}_{\widehat{K}}$\;
  
  Compute $d_i^{(s)}$ for all $i\in[n]$ and form
  $D^{(s)}=\operatorname{diag}(d_1^{(s)},\ldots,d_n^{(s)})$\;
  
  Set
  $S^{(s)}
  =
  \{D^{(s)}+\tau_n\,I_n\}^{-1}
  U_{\widehat K}^{(s)}\Sigma_{\widehat K}^{(s)}$\;
}

Form
$X^{(l)}
=
[S^{(1)},\ldots,S^{(l-1)},S^{(l+1)},\ldots,S^{(L)}]$\;

\KwRet{$X^{(l)}$}
\end{algorithm}

MARC therefore yields nodal covariates $X^{(l)}$ that can be directly incorporated into network-with-covariates methods together with $A^{(l)}$. 
The cross-layer membership heterogeneity is now translated into a mismatch between network and covariates. 
We next address this mismatch when combining $A^{(l)}$ and $X^{(l)}$ for layer-specific community detection.

\subsection{Layer-specific community detection using MARC}
\label{sec-community}
We next combine the covariate matrix $X^{(l)}$ constructed by MARC with the target network $A^{(l)}$ to recover $\pi^{(l)}$.

Recall that cross-layer membership heterogeneity may induce a mismatch between the community structure represented by $X^{(l)}$ and that of $A^{(l)}$. To accommodate this mismatch, we build on the network-adjusted covariate approach of \citet{hu2024network}. 
This approach uses covariates aggregated over network neighbors to represent network information and each node's own covariates to represent covariate information. 
The two components are balanced through a carefully designed weighting scheme, so that the network remains the primary source of community structure while the covariates provide complementary support for extremely sparse nodes.
We adapt this weighting principle to the MARC setting by introducing a node-specific weight determined by the target-layer degree.

Specifically, for target layer $l$, we define
\begin{align}
Y^{(l)}
&=
A^{(l)}X^{(l)}
+
\beta^{(l)}X^{(l)},
\qquad
\beta^{(l)}
=
\operatorname{diag}
\bigl(
\beta_1^{(l)},\ldots,\beta_n^{(l)}
\bigr),
\label{eq:nac}
\end{align}
where
$\beta_i^{(l)}
=
\frac{\bar d^{(l)}/2}
{d_i^{(l)}\log\log(n)/\log n+1}$,
and 
$\bar d^{(l)}
=
\frac{1}{n}\sum_{i=1}^n d_i^{(l)}$.
The weight $\beta_i^{(l)}$ decreases with the target-layer degree $d_i^{(l)}$, so that nodes with fewer observed neighbors place relatively greater weight on their own covariates, whereas nodes with larger degrees rely more heavily on covariates aggregated through the target network.

The behavior of this weight differs sharply across the two sparsity regimes. 
Under our sparsity setting, the relatively sparse nodes determine the order of the average degree, giving $\bar d^{(l)}\asymp\log n$.
For a relatively sparse node with $d_i^{(l)}\gtrsim\log n$, we have $\beta_i^{(l)} \lesssim \frac{\log n}{\log\log n}$, so that the contribution of the node's own covariates is relatively limited compared with neighborhood aggregation. In contrast, for an extremely sparse node with
$d_i^{(l)}
=
o\!\left(\frac{\log n}{\log\log n}\right)$,
we have $\beta_i^{(l)}
=
\{1+o(1)\}\frac{\bar d^{(l)}}{2}
\asymp\log n$.
Thus, each node's own covariates receive substantially greater relative weight when its target-layer neighborhood is extremely sparse. 
For nodes with intermediate degrees, the weight varies continuously between these two scales.

\begin{algorithm}[htbp]
\caption{Layer-specific community detection using MARC}
\label{algo-lasc}
\KwIn{Adjacency matrices $\{A^{(l)}\}_{l=1}^L$, 
spectral dimension $\widehat K$, number of communities $K$}

\textbf{Step 1: Covariate construction.}

Using the MARC construction in Algorithm~\ref{algo-cc}, compute the layerwise blocks $\{S^{(s)}\}_{s=1}^L$\;

\For{$l=1,\ldots,L$}{
  Form
  $X^{(l)}
  =
  [S^{(1)},\ldots,S^{(l-1)},S^{(l+1)},\ldots,S^{(L)}]$\;
}

\textbf{Step 2: Layer-specific community detection.}
\For{$l=1,\ldots,L$}{
  Compute degrees $d_i^{(l)}$ for all $i\in[n]$ and set
  $\bar d^{(l)}=n^{-1}\sum_{i=1}^n d_i^{(l)}$\;
  
  Set
  \[
  \beta^{(l)}
  \gets
  \operatorname{diag}\!\left(
  \frac{\bar d^{(l)}/2}
       {d_1^{(l)}\log\log n/\log n+1},
  \ldots,
  \frac{\bar d^{(l)}/2}
       {d_n^{(l)}\log\log n/\log n+1}
  \right)
  \]
  
  Set
  $Y^{(l)}
  \gets
  A^{(l)}X^{(l)}+\beta^{(l)}X^{(l)}$\;
  
  Compute the $K$ leading left singular vectors of $Y^{(l)}$ and collect them in $\widehat\Xi^{(l)}$\;
  
  Normalize each row of $\widehat\Xi^{(l)}$ to have unit norm,
  and denote the resulting matrix by $\widetilde\Xi^{(l)}$\;
  
  Apply $K$-means clustering to the rows of
  $\widetilde\Xi^{(l)}$ with $K$ clusters to obtain
  $\widehat\pi^{(l)}$\;
}

\KwRet{$\{\widehat\pi^{(l)}\}_{l=1}^L$}
\end{algorithm}
The layerwise spectral blocks $S^{(s)}$ need to be computed only once and can be reused when constructing $X^{(l)}$ for different target layers.
With this reuse, the overall computational complexity of Algorithm~\ref{algo-lasc} is $O(L^2n^2\widehat K)$.
Although our theoretical analysis is developed under the multi-DCSBM, the procedure itself is computed directly from the observed adjacency matrices and does not require fitting the multi-DCSBM.

\section{Theoretical Analysis}
\label{sec-theory}

\subsection{Assumptions and cross-layer complementarity}
To study the theoretical properties of MARC and the community detection approach, we consider the multi-DCSBM in Section~\ref{sec-pre} and specify the assumptions. These assumptions allow us to establish the consistency of MARC covariates and the community recovery guarantees. 
We first impose a cross-layer independence assumption.
\begin{assumption}\label{ass:overall-indep}
Conditional on the community labels \(\{\pi^{(l)}\}_{l=1}^L\), the adjacency matrices \(\{A^{(l)}\}_{l=1}^L\) are mutually independent.
\end{assumption}

We next define a community-level sparsity structure in the multi-DCSBM. As discussed, we allow nodes with extremely sparse connections. 
The sparsity level is shared by nodes within one community. We consider the relatively sparse communities with $E[d_i] \geq c \log n$ and the extremely sparse communities with $E[d_i] \ll \log n$. Thus, the theoretical analysis focuses on the two community-level sparsity regimes, while the procedure itself remains defined for nodes with intermediate degrees. The formal definition is given below. 
\begin{definition}[Community-level sparsity structure]
\label{def:degree-sbm}
Under the multi-DCSBM, let
\(\theta_{\max}=\max_{l\in[L],\,i\in[n]}\theta_i^{(l)}\) and
\(\mathcal I_k^{(l)}=\{i\in[n]:\pi^{(l)}(i)=k\}\).
We say that the model satisfies the community-level sparsity structure if there exists a constant $c_d>0$ such that, for each layer $l$,
there is a layer-specific partition of communities 
$[K]=\mathcal K_D^{(l)}\mathbin{\dot\cup}\mathcal K_S^{(l)}$
satisfying
\begin{itemize}
    \item (Relatively sparse community). If $k \in \mathcal K_D^{(l)}$, then $n\theta_i^{(l)}\theta_{\max} \geq c_d \log n$ for any node $i$ with $\pi^{(l)}(i) = k$;
    \item (Extremely sparse community). If $k \in \mathcal K_S^{(l)}$, then $n\theta_i^{(l)}\theta_{\max} \ll \frac{\log n}{\log \log n}$ for any node $i$ with $\pi^{(l)}(i) = k$.
\end{itemize}
The corresponding relatively sparse and extremely sparse node sets in layer $l$ are
\(\mathcal D^{(l)}=\{i \in [n]: \pi^{(l)}(i) \in \mathcal K_D^{(l)}\}\) and
\(\mathcal S^{(l)}=\{i \in [n]: \pi^{(l)}(i) \in \mathcal K_S^{(l)}\}\),
respectively. Thus,
\([n]=\mathcal D^{(l)}\mathbin{\dot\cup}\mathcal S^{(l)}\).
\end{definition}

For the relatively sparse communities, we use the information from a single target layer. Hence, we need standard assumptions on $\mathcal D^{(l)}$. For the extremely sparse communities, we require sufficient information from other layers. Therefore, we impose Assumptions~\ref{ass:layerwise}--\ref{ass:complementarity} below.   

Let \(K_D^{(l)} = |\mathcal K_D^{(l)}|\) denote the number of relatively sparse communities in layer $l$, and let \(B_D^{(l)} = B^{(l)}_{\mathcal K_D^{(l)}, \mathcal K_D^{(l)}}\) denote the block connection submatrix corresponding to relatively sparse communities. 
We impose the following regularity conditions.
\begin{assumption}
\label{ass:layerwise}
Under the multi-DCSBM, let $\{B^{(l)}\}_{l=1}^{L}$ be constant matrices with each entry in $[0,1]$, and assume the following holds for all $l \in [L]$.
\begin{enumerate}[(i)]
    \item There exists a constant $\rho_s > 0$, so that \(|\mathcal I_k^{(l)}|\geq\rho_s n\)
for every \(k\in[K]\).
\item Every entry of $B^{(l)}$ falls into $[0,1]$ and $\|B^{(l)}\| \leq c_{bu}$ for a constant $c_{bu} > 0$.
\item There exist constants $c_{\theta}, C_{\theta} > 0$, so that \(c_\theta\sqrt{\log n/n}\leq\theta_i^{(l)}
\leq C_\theta\sqrt{\log n/n}\), for every $i \in \mathcal D^{(l)}$. Furthermore, $\mathrm{rank}(B_D^{(l)}) = K_{D}^{(l)}$.
\end{enumerate}
\end{assumption}
For two layers \(l\neq s\), let \(\kappa_{ls}: [K] \to [K]\) denote a permutation achieving the minimum mismatch rate \(q_{ls}\) in Section~\ref{sec-pre}. 
If the minimizer is not unique, we fix one throughout. 
Thus, \(\kappa_{ls}\) maps community labels in layer \(s\) to those in layer \(l\), and \(\kappa_{ls}^{-1}(k)\) denotes the community in layer \(s\) aligned with community \(k\) in layer \(l\). 
\begin{assumption}
\label{ass:complementarity}
For each target layer \(l\), there exists a constant \(c_l\in(0,1]\)
such that
\[
\min_{k\in[K]}
\frac{1}{L-1}
\sum_{s\neq l}
I\left\{
\kappa_{ls}^{-1}(k)\in\mathcal K_D^{(s)}
\right\}
\geq c_l .
\]
\end{assumption}
Assumption~\ref{ass:layerwise}(i) prevents any community from vanishing. Assumption~\ref{ass:layerwise}(ii) is satisfied naturally with a proper scale of $\Theta$. Assumption~\ref{ass:layerwise}(iii) requires that $\theta_i^{(l)}$ for all $i \in \mathcal D^{(l)}$ are of order $\sqrt{\log n/n}$, and the connectivity matrix is full-rank.  
Assumption~\ref{ass:complementarity} requires every community in the target layer to have an aligned relatively sparse counterpart in at least \(c_l(L-1)\) auxiliary layers. It ensures sufficient information from other layers.

\subsection{Theoretical guarantees for MARC covariates}
\label{sec:marc}
We next study the MARC covariates \(X^{(l)}\) constructed from the auxiliary layers in this section. 
To study their properties, we start with an oracle counterpart $\bar{X}^{(l)}$ that delivers target-aligned community information, and then establish the convergence of $X^{(l)}$ to this oracle counterpart $\bar{X}^{(l)}$. Together, these results show that \(X^{(l)}\) is informative in community detection. 

Recall that \(X^{(l)}\) stacks $S^{(s)}$ from auxiliary layers $s \neq l$, so we start with $S^{(s)}$. 
For each source layer \(s\), define
\(\tilde\theta_k^{(s)} = |\mathcal I_k^{(s)}|^{-1/2}\|\theta_{\mathcal I_k^{(s)}}^{(s)}\|\), the scaled $\ell_2$ norm of $\theta^{(s)}$ restricted to community $k$. Recall that $d^{(s)*}_i = \sum_{j\neq i} A_{ij}^{(s)*}$ denotes the population expected degree. 
Construct \(S^{(s)*}\) from the eigenvalue-weighted,
rank-\(K_D^{(s)}\) embedding of
\(A^{(s)*}_{\mathcal D^{(s)},\mathcal D^{(s)}}\)
by multiplying each relatively sparse row \(i\) by
\(\{d^{(s)*}_i  + \tau_n\theta_i^{(s)}/\tilde\theta_{\pi^{(s)}(i)}^{(s)}\}^{-1}\),
and padding with zero rows for extremely sparse nodes and zero columns to obtain an \(n\times\widehat K\) matrix, where \(\widehat K\geq K\). 
Hence, $S^{(s)*}$ is an oracle counterpart of $S^{(s)}$. Furthermore, the $\theta_i^{(s)}$ in the embedding and \(\{d^{(s)*}_i +\tau_n\theta_i^{(s)}/\tilde\theta_{\pi^{(s)}(i)}^{(s)}\}^{-1}\) cancel with each other, so its rows are identical within each source community. 

Our goal is that $\bar{X}^{(l)}$ preserves the target-layer information, so we construct $\bar S^{(s,l)}$ to filter the information of $S^{(s)*}$ aligned with $\pi^{(l)}$ on the target layer $l$. 
Recall that \(\kappa_{ls}^{-1}(\pi^{(l)}(i))\) represents the community in layer $s$ corresponding to the community $\pi^{(l)}(i)$ by the optimal community alignment between two layers. 
Therefore, define row \(i\) of \(\bar S^{(s,l)}\) as the row of \(S^{(s)*}\) corresponding to source community
\(\kappa_{ls}^{-1}(\pi^{(l)}(i))\), which means we refines $S^{(s)*}$ by the community in layer $l$. 
Finally, we stack these blocks to define the label-aligned oracle of $X^{(l)}$, i.e. 
\[
\bar X^{(l)}
=
[\bar S^{(1,l)},\ldots,\bar S^{(l-1,l)},
 \bar S^{(l+1,l)},\ldots,\bar S^{(L,l)}].
\]
The exact formula is provided in the supplementary materials. We then provide the theoretical results about $\bar{X}^{(l)}$ and the deviation $X^{(l)} - \bar{X}^{(l)}$.

\begin{theorem}[Properties of MARC covariates]
\label{thm:marc-covariates}
Suppose Assumptions~\ref{ass:overall-indep}--\ref{ass:complementarity} hold for a multi-DCSBM with community-level sparsity structure. 
Let $\tau_n = c \log n$ for a constant $c > 0$. Fix a target layer \(l\in[L]\). There exist constants
\(c_X, C_X > 0\), so that $\sigma_K(\bar{X}^{(l)})
\geq
\frac{c_X}{1 + \tau_n/\log n}\sqrt{\frac{L-1}{n}}$ and 
\[
\frac{c_X}{1 + \tau_n/\log n}\sqrt{\frac{L-1}{n}}
\leq
\|e_i^\top\bar X^{(l)}\|
\leq
\frac{C_X}{1 + \tau_n/\log n}\sqrt{\frac{L-1}{n}},
\quad i \in [n].
\]
Define the maximum deviation between nodal degree parameter and the community degree parameter across all auxiliary layers $s \neq l$, that 
$\delta_l
:=
\max_{s\ne l}\max_{i\in\mathcal D^{(s)}}
\bigl|
{\theta_i^{(s)}}/
{\widetilde\theta_{\pi^{(s)}(i)}^{(s)}}-1
\bigr|$. Define the average mismatching rate for layer $l$ that \(
\bar q_l=(L-1)^{-1}\sum_{s\neq l}q_{ls}.
\)
Then there exists a block-diagonal orthogonal matrix
\(O^{(-l)}\) and constant $C > 0$ such that with probability \(1-O(1/n)\),
\[
\frac{\|X^{(l)}O^{(-l)}-\bar X^{(l)}\|}
     {\|\bar X^{(l)}\|}
\leq
C\left[
\left(1+\frac{1}{c}\right)
\left\{\frac{1}{\sqrt{\log n}}+\frac{\max_{s \neq l}\max_{i \in \mathcal S^{(s)} }d_i^{(s)*}}{\log n}\right\}
+\frac{c}{1+c}\,\delta_{l}
+\sqrt{\bar q_l}
\right].
\]
Therefore, if the within-community heterogeneity $\delta_l = o(1)$ and the mismatching rate $\bar{q}_l = o(1)$, then the relative error converges to 0. 
\end{theorem}
Theorem~\ref{thm:marc-covariates} first shows the incoherence of the oracle MARC covariates and that it has rank $K$, so the constructed covariates reveal a stable community structure for the target layer. The convergence result then characterizes how well the empirical covariates recover this oracle. The relative spectral error is controlled by a $(\log n)^{-1/2}$ term from the noise in the adjacency matrix, the contribution of sparse communities evaluated by degree scale, the within-community degree heterogeneity $\delta_l$, and the cross-layer membership mismatch \(\bar q_l\).

Under our model, we have $\max_{s \neq l}\max_{i \in \mathcal S^{(s)} }d_i^{(s)*} = o(\log n /\log\log n)$, so the relative error converges to 0 when $\delta_l = o(1)$ and $\bar{q}_l = o(1)$. This model assumption can be relaxed, at the cost of a slower convergence rate. The proof can also be generalized to any choice of $\tau_n$, where $c$ should be replaced by $\tau_n/\log n$. We keep the standard choice of $\tau_n = c\log n$ as in \cite{qin2013regularized}.

\subsection{Weak consistency of layer-specific community detection using MARC covariates}
\label{sec:effect-X}
We now establish the community detection consistency. Recall that the community detection algorithm in Algorithm~\ref{algo-lasc} is to construct $Y^{(l)}$ based on $A^{(l)}$ and $X^{(l)}$, and then apply spectral clustering on $Y^{(l)}$. We propose an oracle counterpart $\bar{Y}^{(l)}$ first, and then demonstrate the weak consistency result. 

Fix a target layer $l \in [L]$. Let $I_{\mathcal D^{(l)}}$ and $I_{\mathcal S^{(l)}}$ be the diagonal selector for nodes in the relatively sparse set $\mathcal D^{(l)}$ and extremely sparse set $\mathcal S^{(l)}$, respectively. 
Define the oracle matrix $\bar{A}^{(l)} =I_{\mathcal D^{(l)}}\, A^{(l)*} I_{\mathcal D^{(l)}}$, which retains the connections among relatively sparse nodes $\mathcal D^{(l)}$. The information of extremely sparse nodes $\mathcal S^{(l)}$ is contained in $I_{\mathcal S^{(l)}} \bar{X}^{(l)}$. Accordingly, let $\beta^{(l)*}$ be the oracle weight matrix (see supplementary materials for definition) and define the oracle matrix
\begin{equation}
\label{eq:Y-oracle-short}
\bar{Y}^{(l)}
=
\bar{A}^{(l)}\bar{X}^{(l)}
+
I_{\mathcal S^{(l)}}
\beta^{(l)*}\bar{X}^{(l)}.
\end{equation}
Recall that $\bar{X}^{(l)}$ contains cross-layer information aligned with the target layer membership. Therefore, the first term in \eqref{eq:Y-oracle-short} aggregates cross-layer information through target-layer neighbours, while the second term focuses on the covariate information of the individual node. With the balancing scheme $\beta^{(l)*}$, we have the following lemma. 
\begin{lemma}[Oracle geometry] \label{lemma:oracle}
Assume $\widehat K\ge K$.
Let
$\bar{Y}^{(l)}
=
\bar{\Xi}^{(l)}
\bar{\Lambda}^{(l)}
\bar{V}^{(l)\top}$
be a rank-$K$ singular value decomposition.
Then the $i$-th row of $\bar{\Xi}^{(l)}$ has a community-specific direction:
\[
e_i^\top\bar{\Xi}^{(l)}
=
c_i\,e_{\pi^{(l)}(i)}^\top Q^{(l)},
\quad
\text{for some } c_i>0
\text{ and a nonsingular } Q^{(l)}\in\mathbb R^{K\times K}.
\]
Hence, row normalization of $\bar{\Xi}^{(l)}$ yields exact recovery of
$\pi^{(l)}$, up to a label permutation.
\end{lemma}

Lemma~\ref{lemma:oracle} shows that spectral clustering on the oracle matrix $\bar{Y}^{(l)}$, followed by row normalization, exactly recovers $\pi^{(l)}$ up to a label permutation. This oracle case establishes the basic idea underlying our community detection procedure. We next show that the empirical version $Y^{(l)}$ can achieve weak consistency.
\begin{theorem}[Weak consistency]
\label{the:weak}
Suppose Assumptions~\ref{ass:overall-indep}--\ref{ass:complementarity} hold with fixed $K$ and $L$. 
Fix a target layer $l$ with $K_D^{(l)}\geq 1$. 
Let $\widehat K=K$ in the MARC construction, and choose $\tau_n=c\log n$ for a sufficiently small constant $c>0$ depending only on the model constants. 
Then there exists a constant $C>0$ such that
with probability $1-o(1)$,
\[
\min_{\kappa\in\mathfrak S_K}\frac1n
\big|\{i\in[n]:
\kappa(\widehat\pi^{(l)}(i))\ne\pi^{(l)}(i)\}\big|
\leq
C\left\{
\bar q_l+\frac1{(\log n)^2}
\right\}.
\]
Therefore, when $\bar{q}_l = o(1)$, the misclustering error rate of Algorithm~\ref{algo-lasc} goes to 0.  
\end{theorem}

Theorem~\ref{the:weak} establishes the weak consistency of Algorithm~\ref{algo-lasc}. 
The misclustering error rate consists of two leading components: cross-layer membership mismatch contributes \(\bar q_l\), and the spectral estimation error contributes \(O((\log n)^{-2})\). 
Hence, weak consistency holds when $\bar{q}_l \to 0$. 

The condition $\bar{q}_l \to 0$ is comparable to those obtained under more restrictive models. 
\citet{chen2022global} consider the case $K = 2$ with a common label $\pi$. The layer-specific label $\pi^{(l)}(i)$ agrees with $\pi(i)$ with probability $1 - \rho$ and is flipped with probability $\rho$, giving $\bar q_l = 2\rho(1-\rho)+o_{\mathbb P}(1)$. Their community detection error is of order $\rho$. 
Under this model, \citet{lei2023bias} yield the same $O(\rho)$ dependence. 
\citet{agterberg2025joint} consider a more general model; when specialized to this two-community setting, their method requires $\rho+O_{\mathbb P}(\sqrt{\log n/L})$ to vanish. Our method considers a more flexible multi-DCSBM that allows general layer-specific memberships and extremely sparse communities, while weak consistency still requires only $\bar q_l\to0$ as the membership-mismatch condition.

We finally note an important difference between Theorem~\ref{thm:marc-covariates} and Theorem~\ref{the:weak}. The condition $\delta_l = o(1)$ is required for covariate consistency but not necessary for weak consistency. The reason is that Algorithm~\ref{algo-lasc} only requires the community centers to stay apart. The oracle covariates $\bar{X}^{(l)}$ have centers separated at the $n^{-1/2}$ scale. The heterogeneity $\delta_l$ introduces an error of $cn^{-1/2}$ between $X^{(l)}$ and $\bar{X}^{(l)}$, where $c = \tau_n/\log n$, under the row-wise norm. When $c$ is sufficiently small, the community centers are still well separated, so $\delta_l$ need not vanish for weak consistency.

\subsection{Strong consistency with weak-community cores}
\label{sec:sc-main}
Exact recovery requires uniform row-wise control of the estimated embedding and therefore stronger
conditions than those required for weak consistency. To control the localization of the $K - K_D^{(l)}$ weak spectral directions, we have additional assumptions on the node degrees of extremely sparse communities. Let $\mathcal H_k^{(l)}$ be a core in community $k$ with higher degrees. The proportion of the core $\mathcal H_k^{(l)}$, $\gamma_n$, can be a constant or decay with $n$. The degree heterogeneity parameter depends on $\gamma_n$. 
Detailed assumptions are as follows.
\begin{assumption}
\label{ass:sc-cores}
Under the multi-DCSBM, assume the following.
\begin{enumerate}
\item[(i)] All matrices $B^{(l)}$ are positive definite, with
$\lambda_{\min}(B^{(l)})\geq c_{bl}>0$, $l\in[L]$.
\item[(ii)] (Core structure of extremely sparse communities)
There exist $0<\gamma_n\leq1$ and $\kappa_n$ satisfying
$(\log n)^{2/3}\ll\kappa_n\ll(\log n)^{1-\epsilon}\gamma_n$
for some constant $\epsilon>0$, such that every extremely sparse community $k$ has a core $\mathcal H_k^{(l)}$, with the proportion $c \gamma_n \leq {|\mathcal H_k^{(l)}|}/{|\mathcal I_k^{(l)}|}
\leq  C\gamma_n$ for constants $c, C > 0$, and
\begin{equation}
\label{eq:sc-core-rates}
\theta_i^{(l)}
\asymp
\sqrt{\frac{\kappa_n}{n\gamma_n}},
\quad i\in\mathcal H_k^{(l)};
\qquad
\theta_i^{(l)}
\asymp
\frac{1}{\sqrt{n\log n}},
\quad i\in\mathcal I_k^{(l)}\backslash \mathcal H_k^{(l)}.
\end{equation}

\end{enumerate}
\end{assumption}

Assumption~\ref{ass:sc-cores} introduces additional assumptions for uniform error control over all nodes. Assumption~\ref{ass:sc-cores}(i) requires the whole matrix $B^{(l)}$ to be uniformly positive definite, rather than only its relatively sparse block. Assumption~\ref{ass:sc-cores}~(ii) specifies the core structure in extremely sparse communities. For each extremely sparse community, there is a core $\mathcal H_k^{(l)}$ whose nodes have within-community expected degrees of order $\kappa_n$, while the remaining nodes have much lower expected degrees of order $O(1)$. The core proportion $\gamma_n$ may be constant or vanish with $n$.

Exact recovery requires uniform control of the row-normalized target embedding for each node. Therefore, we have to control the mismatching rate for each extremely sparse node. 
Define the node-specific mismatch
\[
\zeta^{(l)}(i)
=
\frac{1}{L-1}\sum_{s\ne l}
I\{\kappa_{ls}(\pi^{(s)}(i))\ne\pi^{(l)}(i)\},
\]
so that the average $n^{-1}\sum_i\zeta^{(l)}(i)=\bar q_l$.
The following theorem establishes exact recovery when the average
mismatch $\bar{q}_l$ is sufficiently small, and the node-specific mismatch rates $\zeta^{(l)}(i)$ for extremely sparse nodes are sufficiently small.
\begin{theorem}[Strong consistency]
\label{thm:sc-strong}
Under the multi-DCSBM, suppose Assumptions~\ref{ass:overall-indep}--\ref{ass:sc-cores} hold. Fix $K, L \geq 2$ and $\widehat{K} = K$. For a fixed layer $l \in [L]$ with $K_{D}^{(l)} \geq 1$, let $\hat{\pi}^{(l)}$ be the estimated membership vector from Algorithm~\ref{algo-lasc}.
There is a constant $C\geq1$ such that,
if
\begin{equation}
\label{eq:sc-recovery-margin}
 \limsup_{n\to\infty}
 C\left\{
   \frac1{c_\theta}+\frac{1+c}{c_\theta^2}
   +\sqrt{\bar q_l}
   +\max_{i\in\mathcal S^{(l)}}\zeta^{(l)}(i)
 \right\}<\frac{\sqrt2}{4},
\end{equation}
then, with probability $1 - o(1)$, $\min_{\kappa\in\mathfrak S_K}\sum_{i=1}^nI\{\kappa(\widehat\pi^{(l)}(i))\ne\pi^{(l)}(i)\}=0$.
\end{theorem}
Unlike the weak-consistency result in Theorem~\ref{the:weak}, strong consistency in Theorem~\ref{thm:sc-strong} does not require the average mismatch $\bar q_l$ to vanish. 
Under the additional assumptions, the row-wise perturbation induced by cross-layer mismatch is of order $\zeta^{(l)}(i)n^{-1/2}$, where $\zeta^{(l)}(i)$ denotes its node-specific mismatch rate. 
Since the oracle community centers are separated at the same $n^{-1/2}$ scale, the mismatch only needs to be sufficiently small,
rather than vanishing, so that the empirical rows remain within the correct clustering margin. 
The additional nodewise mismatch condition is imposed only on extremely sparse target nodes $\mathcal S^{(l)}$, whose recovery relies more heavily on auxiliary covariates. Relatively sparse target nodes $\mathcal D^{(l)}$ benefit from neighborhood aggregation and do not require an analogous uniform mismatch condition. The guarantee applies to the full procedure with the weak spectral columns retained.

The basic proof idea is to separate the retained source spectral directions $X^{(l)}$ into reliable and weak blocks and trace their contributions to $Y^{(l)}$. Define
\begin{equation}
\label{eq:sc-column-split}
Z=(A^{(l)}+\beta^{(l)})
  [S^{(s)}_{:,1:K_D^{(s)}}]_{s\ne l},
\qquad
F=(A^{(l)}+\beta^{(l)})
  [S^{(s)}_{:,K_D^{(s)}+1:K}]_{s\ne l}.
\end{equation}
Then $Y^{(l)}=[Z,F]$ up to a column permutation. 
Here $Z$ collects the reliable directions associated with relatively sparse communities, while $F$ contains the remaining weak directions. Exact recovery requires $Z$ to retain the full target community structure and the contribution of $F$ to the target embedding to be sufficiently small.

For $Z$, we use the row-wise analysis of \citet{hu2024network} to control the error in the reliable source directions. The additional difficulty is that each auxiliary layer contributes only $K_{\mathcal{D}}^{(s)}$ reliable directions,  which may be fewer than $K$. We therefore need to show that these directions jointly retain the full target community structure. Cross-layer complementarity in Assumption~\ref{ass:complementarity}, together with $\lambda_{\min}(B^{(l)}) \geq c_{bl} > 0$ in Assumption~\ref{ass:sc-cores}, ensures that $Z$ retains a stable rank-$K$ signal, even when no individual auxiliary layer provides all $K$ reliable directions.

The matrix $F$ contains the weak directions, which may arise from extremely sparse communities or from sampling noise in relatively sparse communities. These directions can be localized, so their small eigenvalues alone do not guarantee a sufficiently small row-wise norm of $F$. 
However, exact recovery only requires their contribution to the leading target embedding to be small. We therefore consider
\begin{equation}
\label{eq:sc-weak-contribution}
\Delta_F
=
F\widehat V_F^{(l)}(\widehat\Lambda^{(l)})^{-1}
=
FF^\top\widehat\Xi^{(l)}(\widehat\Lambda^{(l)})^{-2},
\end{equation}
where $\widehat V_F^{(l)}$ is the block of the leading right singular
vectors of $Y^{(l)}=[Z,F]$ corresponding to $F$.

Assumption~\ref{ass:sc-cores}(ii) gives one degree configuration under which this projected contribution can be controlled. Under Assumption~\ref{ass:sc-cores}(ii),
$\|FF^\top\|_\infty
=
O_{\mathbb P}(\kappa_n\log n)
=
o_{\mathbb P}\{(\log n)^2\}$. Together with the row-wise and singular-value bounds for $Z$, it gives
$\|\Delta_F\|_{2\to\infty}=o_{\mathbb P}(n^{-1/2})$.
The contribution is negligible under the node-specific balancing for extremely sparse target nodes. Here, the core structure in Assumption~\ref{ass:sc-cores}(ii) is sufficient rather than necessary. Other degree configurations, such as uniformly sparse communities, also ensure the same projected-error control.
Combining these bounds with the membership-mismatch and cluster-center errors yields Theorem~\ref{thm:sc-strong}.

\section{Global consensus from layer-specific communities}\label{sec-global}

\subsection{Label alignment and weighted consensus}
\label{sec-consensus-method}
By the layer-specific community detection approach in Algorithm~\ref{algo-lasc}, we have community memberships $\{\widehat\pi^{(l)}\}_{l\in[L]}$ for each layer.
Although these memberships need not coincide, a global consensus $\widehat\pi$ may provide a useful summary of their common structure. 
In this section, we provide such a summary by aligning the community labels across layers and forming a weighted combination of the estimated
membership matrices.
This is an optional step performed after layer-specific recovery, which does not modify the estimated memberships within each layer.

Let $\widehat\Pi^{(l)}\in\{0,1\}^{n\times K}$ denote the membership
matrix associated with $\widehat\pi^{(l)}$.
For each layer $l$, let $R^{(l)}\in\mathcal O(K)$ be a permutation
matrix, where
$\mathcal O(K)=\{R\in\{0,1\}^{K\times K}:R^\top R=I_K,\,
R1_K=1_K\}$.
The matrix $R^{(l)}$ changes only the community labels, rather than
the partition itself.
We also introduce layer weights
$\xi\in\Delta_L=\{\xi\in[0,\infty)^L:\sum_{l=1}^L\xi_l=1\}$.
We seek the alignment matrices and weights by maximizing
\begin{equation}
\max_{\substack{\xi\in\Delta_L,
                 R^{(l)}\in\mathcal O(K),\,l\in[L]}}
\sum_{1\leq l<s\leq L}
2\xi_l\xi_s
\operatorname{tr}\left\{
R^{(l)\top}\widehat\Pi^{(l)\top}
\widehat\Pi^{(s)}R^{(s)}
\right\}.
\label{eq:consensus-objective}
\end{equation}
The trace counts the nodes assigned to the same community in the
two aligned partitions.
Thus, the objective rewards weighted cross-layer agreement:
the permutation matrices establish label correspondences, while
the weights determine each layer's contribution to the consensus.
Only agreement between distinct layers enters the objective.

For computation, we alternate between updating the alignment
matrices and the layer weights.
With $\xi$ and $\{R^{(s)}\}_{s\neq l}$ fixed, the alignment
subproblem for layer $l$ is
\begin{equation}
\max_{R\in\mathcal O(K)}
\operatorname{tr}\biggl\{
R^\top\widehat\Pi^{(l)\top}
\sum_{s\neq l}\xi_s\widehat\Pi^{(s)}R^{(s)}
\biggr\}.
\label{eq:consensus-alignment}
\end{equation}
The conditional update of $R^{(l)}$ is a linear assignment problem. 
In Algorithm~\ref{alg:consensus}, we obtain an approximate solution by solving an orthogonal relaxation and projecting the resulting
matrix onto $\mathcal O(K)$. 
With the alignment matrices fixed, the objective in
\eqref{eq:consensus-objective} is quadratic in $\xi$ and has an
explicit gradient.
We update the weights by projected gradient ascent over $\Delta_L$.
Algorithm~\ref{alg:consensus} summarizes the alternating procedure.

Let $\widehat\xi$ and $\{\widehat R^{(l)}\}_{l\in[L]}$ denote
the resulting weights and alignment matrices.
The consensus matrix and labels are
\begin{equation}
\widehat H
=
\sum_{l=1}^L
\widehat\xi_l\widehat\Pi^{(l)}\widehat R^{(l)},
\qquad
\widehat\pi(i)
\in
\operatorname*{arg\,max}_{k\in[K]}
\widehat H_{ik},
\quad i\in[n].
\label{eq:consensus-output}
\end{equation}
Each entry $\widehat H_{ik}$ records the total weight of layers
assigning node $i$ to community $k$ after alignment.
The final assignment therefore gives a weighted consensus of the
layer-specific memberships, rather than requiring those
memberships to be identical.

\begin{algorithm}[t]
\caption{Label alignment and weighted consensus}
\label{alg:consensus}

\KwIn{Membership matrices $\{\widehat\Pi^{(l)}\}_{l=1}^L$,
step size $\eta>0$, tolerance $\varepsilon>0$,
and maximum iterations $T_{\max}$}

Initialize $\xi\leftarrow 1_L/L$ and
$R^{(l)}\leftarrow I_K$ for $l\in[L]$\;

\For{$t=1,\ldots,T_{\max}$}{
    Update $\{R^{(l)}\}_{l=1}^L$ simultaneously using
    orthogonal relaxations of \eqref{eq:consensus-alignment},
    followed by projection onto $\mathcal O(K)$\;

    Update $\xi$ by projected gradient ascent for
    \eqref{eq:consensus-objective} over $\Delta_L$
    with step size $\eta$\;

    Stop if the relative change in the objective
    \eqref{eq:consensus-objective} is below $\varepsilon$\;
}

Set $\widehat\xi\leftarrow\xi$ and
$\widehat R^{(l)}\leftarrow R^{(l)}$ for $l\in[L]$\;
Compute $\widehat H$ and $\widehat\pi$
using \eqref{eq:consensus-output}\;

\KwRet{$\widehat H,\widehat\pi$}
\end{algorithm}

\subsection{Recovery of an approximately common partition}
We now discuss the theoretical guarantee of the consensus construction in Section~\ref{sec-consensus-method}. To start with, let $\pi^0:[n]\to[K]$ denote a reference partition, with membership matrix $\Pi^0\in\{0,1\}^{n\times K}$. It is introduced only for the consensus analysis and is not assumed in the multilayer model. 

We then describe how the layer-specific labels $\pi^{(l)}$ differ from $\pi^0$.
For each layer $l\in[L]$, let $R^{(l)*}\in{\cal O}(K)$ denote an oracle permutation aligning $\pi^{(l)}$ with $\pi^0$, and write $\widetilde\Pi^{(l)}=\Pi^{(l)}R^{(l)*}$.
Let $\xi^*\in\Delta_L$ denote the population-optimal consensus
weights. Define the overall deviation of the layer-specific memberships from $\pi^0$ as
\[
\eta
=
\max_{k\in[K]}
\frac{
\sum_{l=1}^L \xi_l^*
\big|
\{i:\pi^0(i)=k,\,
      \widetilde\pi^{(l)}(i)\neq k\}
\big|
}{
\big|
\{i:\pi^0(i)=k\}
\big|
}.
\]
Thus, $\eta$ measures the largest weighted discrepancy between a reference community and its aligned layer-specific memberships.
A small value of $\eta$ means that the reference partition $\pi^0$ provides a good weighted summary.

For this analysis, we assume that the layer-specific memberships are recovered exactly up to label permutations. It is satisfied when the exact-recovery conditions in Theorem~\ref{thm:sc-strong} hold for every layer. 
We consider a split-sample version of the consensus construction. Randomly partition $[n]={\cal I}_1\cup{\cal I}_2$, where $n_j=|{\cal I}_j|\asymp n$. The alignment matrices and consensus weights are estimated using ${\cal I}_1$, and the resulting consensus labels are evaluated on ${\cal I}_2$. Theoretical results below are for the global maximizer in Section~\ref{sec-consensus-method}.
\begin{theorem}[Recovery of an approximately common partition]
\label{thm:global-align-revised}
Suppose that $\pi^{(l)}$ are recovered exactly up to label permutations and that identification conditions in the supplementary material hold.
Then, with probability $1-o(1)$,
\[
\frac{1}{n_2}
\min_{\kappa\in\mathfrak S_K}
\big|
\{i\in{\cal I}_2:
\kappa\{\widehat\pi(i)\}\neq\pi^0(i)\}
\big|
\leq
\frac{2\eta}{1-2r_{\xi,n,L}}
+
\sqrt{\frac{\log n}{2n_2}},
\]
where $r_{\xi,n,L}$ controls the deviation between the estimated consensus weights and their population counterparts. Consequently,
if $r_{\xi,n,L}=o(1)$ and $\eta=o(1)$, then the error rate goes to 0
with probability $1-o(1)$.
\end{theorem}

Theorem~\ref{thm:global-align-revised} provides a statistical interpretation of the consensus when an approximately common partition $\pi^0$ exists.
The error is governed by two quantities: the reference partition quality $\eta$ and the consensus weight estimation error $r_{\xi,n,L}$. 
In particular, when the layer-specific memberships become increasingly compatible with $\pi^0$, the consensus consistently recovers that partition.
When no such approximately common partition exists, the construction in Section~\ref{sec-consensus-method} remains well-defined as a weighted summary of the layer-specific memberships.

\section{Simulation}
\label{sec-simu}

\subsection{Covariate-induced multilayer networks}
We first consider multilayer networks obtained by thresholding similarities between covariates from different modalities \citep{wang2021mogonet}. For node $i$, we generate covariates $Z_i^{(l)}\in\mathbb R^p$ and form a cosine-similarity matrix $C^{(l)}$. The network $A^{(l)}$ is constructed by retaining the largest $80n/2$ upper-triangular similarities, corresponding to approximately the top $8\%$ when $n=1000$.

For community labels, we first generate uniform global labels $\pi(i)\in[K]$ independently, together with global community mean vectors from independent Gaussian directions, normalized to have norm $\sigma_M\sqrt p$. Conditional on the global labels, the layer-specific labels $\pi^{(l)}$ are generated independently across layers: for a node with global label $\pi(i)=k$, the label $k$ is retained in layer $l$ with probability $1-q_l$, while with probability $q_l$ it is reassigned uniformly to one of the other $K-1$ communities. Hence, $q_l$ controls the heterogeneity of layer $l$ from the global membership, while the pairwise mismatch $q_{ls}$ is determined by the generated layer-specific labels. Larger transition probabilities correspond to larger pairwise mismatches. For each node $i$, we randomly select $\mathcal I_i\subset[L]$ with $|\mathcal I_i|=\lfloor0.5L\rfloor$. Given $\pi^{(l)}(i)=k$, we generate the covariate $Z_i^{(l)}\sim\mathcal N(0.35\mu_k^{(0)},I_p)$ for $l\in\mathcal I_i$ and $Z_i^{(l)}\sim\mathcal N(\mu_k^{(0)},I_p)$ otherwise. Hence, each node has a weaker covariate signal in approximately half of the layers.

We consider three sets of comparison methods: 
(i) methods targeting layer-specific memberships, including spectral clustering (SPEC; \citealp{chauhan2009spectral}), spectral clustering on ratios-of-eigenvectors (SCORE; \citealp{jin2015fast}), and the layer-specific version of Tucker decomposition with integrated SVD transformation (TWIST-L; \citealp{jing2021community}); 
(ii) global multilayer methods imposing a common membership, including degree-corrected multiple adjacency spectral embedding (DCMASE; \citealp{agterberg2025joint}) and the global version of TWIST (TWIST-G; \citealp{jing2021community}); and 
(iii) network-and-covariate methods, including covariate-assisted spectral clustering (CASC; \citealp{binkiewicz2017covariate}) using the MARC covariates and the proposed community-detection procedure. 
SPEC and SCORE are applied separately to each layer, whereas TWIST-L jointly uses the multilayer network to estimate layer-specific memberships.

We take $n=1000$, $p=20$, $K=4$, $\sigma_M=3.5$, and $L\in\{6,8,\ldots,12\}$. To create substantial cross-layer heterogeneity while allowing the later layers to become progressively more compatible, we set $q_l=0.35$ for $l=1,\ldots,4$ and $q_l=0.20(0.60)^{l-5}$ for $l\ge5$. Thus, the first four layers remain deliberately heterogeneous and difficult, whereas newly added layers become increasingly aligned with the global membership as $L$ increases. Consequently, the later auxiliary layers become progressively more compatible with the first four target layers as $L$ increases. 
We report the average misclustering rate across all layers and, separately, over the first four difficult layers, which allows us to track the same target layers as $L$ increases. Table~\ref{tab:exp1} summarizes the two misclustering rates over 100 repetitions.

\begin{table}[htbp]
\begingroup
\renewcommand{\baselinestretch}{1}\selectfont
\centering
\footnotesize
\setlength{\tabcolsep}{5.2pt}
\renewcommand{\arraystretch}{1.05}
\begin{tabular}{cc cc ccc cc}
\toprule
& & \multicolumn{2}{c}{Single-layer}
& \multicolumn{3}{c}{Multilayer}
& \multicolumn{2}{c}{Network--covariate} \\
\cmidrule(lr){3-4}\cmidrule(lr){5-7}\cmidrule(lr){8-9}
\(L\) & Layers
& SPEC & SCORE & TWIST-L & TWIST-G & DCMASE & CASC & Proposed \\
\midrule

6  & All     & 0.301 & 0.298 & 0.298 & 0.406 & 0.353 & 0.330 & \textbf{0.252} \\
   & First 4 & 0.292 & 0.287 & 0.310 & 0.475 & 0.452 & 0.351 & \textbf{0.267} \\
\addlinespace[2pt]

8  & All     & 0.302 & 0.292 & 0.279 & 0.306 & 0.262 & 0.282 & \textbf{0.208} \\
   & First 4 & 0.297 & 0.293 & 0.301 & 0.417 & 0.371 & 0.294 & \textbf{0.226} \\
\addlinespace[2pt]

10 & All     & 0.293 & 0.286 & 0.295 & 0.204 & 0.223 & 0.198 & \textbf{0.161} \\
   & First 4 & 0.289 & 0.275 & 0.291 & 0.373 & 0.357 & 0.217 & \textbf{0.176} \\
\addlinespace[2pt]

12 & All     & 0.296 & 0.287 & 0.278 & 0.176 & 0.171 & 0.165 & \textbf{0.143} \\
   & First 4 & 0.301 & 0.293 & 0.295 & 0.324 & 0.347 & 0.178 & \textbf{0.160} \\
\bottomrule
\end{tabular}
\caption{Average per-layer misclustering rates for the covariate-induced multilayer networks.
For each \(L\), the first row averages over all layers and the second over the first four difficult layers.
CASC uses the MARC covariates.}
\label{tab:exp1}
\endgroup
\end{table}
Table~\ref{tab:exp1} shows that our proposed community detection procedure achieves the lowest misclustering rate in all settings. For the single-layer methods, SPEC and SCORE, both the overall error and the error over the first four difficult layers remain nearly unchanged as $L$ increases, since they use only the target network. Among the multilayer methods, TWIST-L also shows little improvement, reflecting its greater emphasis on layer-specific memberships. For TWIST-G and DCMASE, the error rates decrease as $L$ increases, especially on the overall misclustering error rate. This indicates that these multilayer methods benefit from borrowing across layers, but they do not adapt the borrowed information sufficiently to the heterogeneity of a particular target layer.

In contrast, both CASC with the MARC covariates and our proposed procedure improve substantially with $L$, both overall and on the first four difficult layers. 
This demonstrates that the MARC covariates effectively extract useful auxiliary information for target-layer community detection, and that converting a heterogeneous multilayer network into a network-with-covariates problem provides an effective way to accommodate cross-layer membership heterogeneity. Our proposed procedure further improves over CASC because its node-specific balancing scheme better accommodates extremely sparse nodes.

\subsection{Heterogeneous multi-DCSBM}
We examine the effect of cross-layer membership heterogeneity under the multi-DCSBM. Consider $n=1000$ nodes, $L=4$ layers, and $K=4$ communities in each layer. 
We generate $\bpi^{(l)}$ as in Experiment 1, using a common transition parameter $q$ across all layers, i.e., $q_1=\cdots=q_L=q$, with $q\in\{0,0.1,0.2,\ldots,1\}$. 
Under this construction, the realized mismatch matrix $Q$ varies across repetitions, while $q_{ls}$ is centered around $2q-Kq^2/(K-1)$ for $l \neq s$. Accordingly, $\bar q_l$ follows the same population pattern. For $K=4$, the population mismatch first increases as $q$ increases, reaches its maximum at $q=0.75$ and then decreases. 

For each layer $l\in[L]$, we randomly select two communities and generate
$\theta_i^{(l)}\sim\mathrm{Unif}(0.10,0.50)$ independently for nodes in these communities. Randomly select another community and generate
$\theta_i^{(l)}\sim\mathrm{Unif}(0.05,0.20)$. The remaining community has $\theta_i^{(l)}\sim\mathrm{Unif}(0.15,0.25)$.
Hence, the latter two communities are relatively sparse. Define the block matrices
\begin{align*}
&B^{(1)}=0.9 I_K+0.1\,1_K1_K^\top,\quad\quad
B^{(2)}=0.7 I_K+0.3\,1_K1_K^\top,\\
&B^{(3)}=0.5 I_K+0.05\,(1_K1_K^\top-I_K),\quad
B^{(4)}=0.05 I_K+0.5\,(1_K1_K^\top-I_K).
\end{align*}
Layers $1$--$3$ are assortative and layer~$4$ is disassortative. Moreover, layer $3$ is sparser than layers $1$--$2$. For nodes $i,j$ with $\pi^{(l)}(i) = k_1$ and $\pi^{(l)}(j) = k_2$, we generate 
$A^{(l)}_{ij}\sim\mathrm{Bernoulli}
(\theta_i^{(l)}\theta_j^{(l)}
B^{(l)}_{k_1, k_2})$.

\begin{figure}[htbp]
    \centering
    \includegraphics[width=\textwidth]{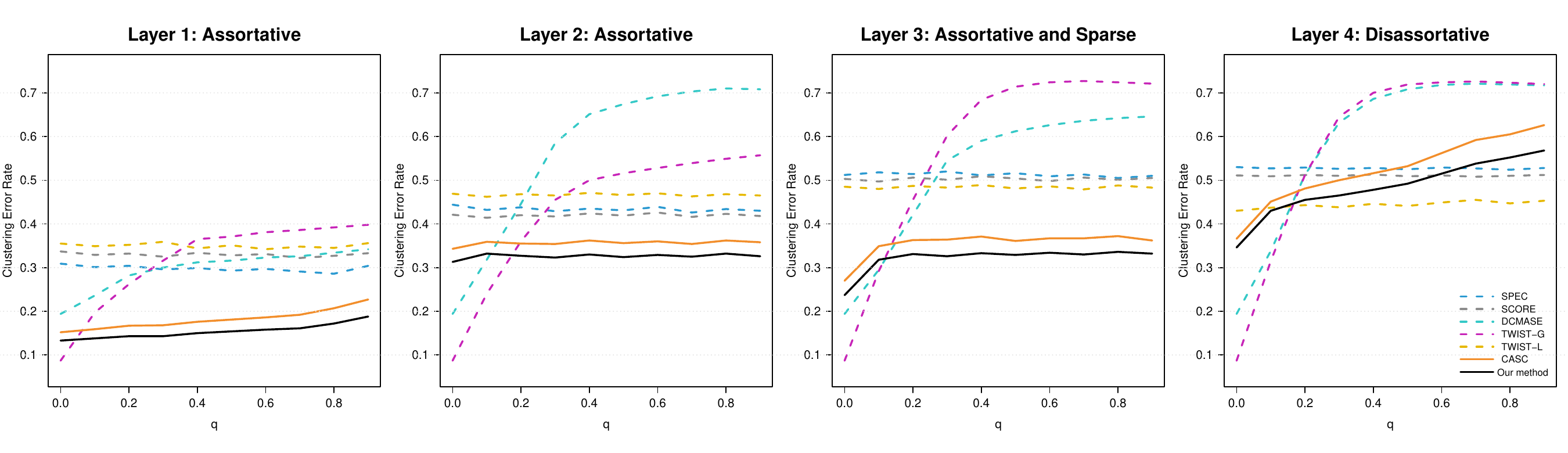}
    \caption{Layer-wise misclustering rate versus membership transition parameter $q$.}
    \label{fig:layer}
\end{figure}

We consider both layer-specific and global misclustering errors over
$q\in\{0,0.1,\ldots,1\}$. 
For layer-specific community detection, we compare with the same methods as in Experiment~1 and summarize the average misclustering error for each layer over 100 repetitions in Figure~\ref{fig:layer}. 
For global community detection, we apply the consensus label-alignment method in Section~\ref{sec-global} to obtain a global membership estimate and evaluate its misclustering error relative to $\pi$. Among the comparison methods, only TWIST-G and DCMASE provide global membership estimates and are therefore included. We additionally consider $\widehat K=K-1$ and $\widehat K=K+2$, besides the default $\widehat K=K$, to examine the effect of the auxiliary spectral dimension. The corresponding results over 100 repetitions are reported in Table~\ref{tab:align_q1}.

Figure~\ref{fig:layer} shows that layer-specific community detection is challenging under sparsity and mixed assortative structures, with misclustering rates as high as 0.7. Our proposed procedure achieves the lowest error in most scenarios, particularly in Layers~1--3 when $q\geq0.2$. CASC with our MARC covariates is the second-best method in these regimes, demonstrating the effectiveness of the MARC covariates and our network-with-covariates representation.
Methods targeting layer-specific memberships, including SPEC, SCORE, and TWIST-L, maintain relatively high error rates across $q$, indicating the need to borrow information from other layers. 
In contrast, TWIST-G and DCMASE perform better when memberships are similar ($q<0.2$), but this advantage reverses once $q>0.4$. Our procedure avoids this tradeoff by borrowing auxiliary information while adapting to membership heterogeneity. 

At the layer level, our proposed procedure performs best in Layers~1--3, with the largest gain in the sparse Layer~3. TWIST-L is relatively weak in Layers~1--2 but becomes the strongest competitor in the sparse and disassortative Layer~4, reflecting its greater emphasis on layer-specific structure. Our procedure also achieves one of the lowest error rates in Layer~4.

Table~\ref{tab:align_q1} summarizes the global membership results. As $q$ increases, all methods have a larger error rate because the layer-specific memberships deviate further from the global membership $\pi$. The proposed consensus alignment consistently outperforms DCMASE and TWIST-G. For $q>0.5$, the error rates remain at similar levels, so the full results are deferred to the supplementary materials. The results are insensitive to moderate misspecification of the auxiliary spectral dimension $\hat K$: over-specification has little effect, while under-specification leads to only a modest increase in error.
\begin{table}[htbp]
\begingroup
\renewcommand{\baselinestretch}{1}\selectfont 
\centering
\footnotesize
\setlength{\tabcolsep}{7pt}
\renewcommand{\arraystretch}{1.05}
\begin{tabular}{c cc ccc}
\toprule
 & \multicolumn{2}{c}{Global baselines} 
 & \multicolumn{3}{c}{Proposed consensus alignment} \\
\cmidrule(lr){2-3}\cmidrule(lr){4-6}
\(q\) & DCMASE & TWIST-G 
& \(\hat K=K\) & \(\hat K=K-1\) & \(\hat K=K+2\) \\
\midrule
0.1 & 0.2786 & 0.3245 & 0.2406 & 0.2504 & 0.2426 \\
0.2 & 0.4324 & 0.4970 & 0.3372 & 0.3504 & 0.3366 \\
0.3 & 0.5088 & 0.5520 & 0.4336 & 0.4588 & 0.4321 \\
0.4 & 0.5658 & 0.6072 & 0.4954 & 0.5058 & 0.4924 \\
0.5 & 0.6250 & 0.6433 & 0.5725 & 0.6000 & 0.5888 \\
\bottomrule
\end{tabular}
\caption{Misclustering error relative to the global membership $\bpi$ under varying $q$ and auxiliary spectral dimension $\hat K$.}\label{tab:align_q1}
\endgroup
\end{table}

We further conduct two experiments by varying the number of communities across layers and introducing cross-layer edge dependence to examine robustness. Across both settings, the proposed procedure yields lower error rates than the competing methods. The results are deferred to the supplementary materials.

\section{Food Trading Networks}
\label{sec-real}
We analyze global food trading networks compiled by \citet{de2015structural} (available at \url{http://www.fao.org}). Each layer corresponds to a product traded in 2010, with nodes representing countries and edges representing bilateral trade links. Following \citet{jing2021community}, we remove edges with weight below 8 and exclude layers whose largest connected component has fewer than 150 nodes. Integrating the largest connected components across the remaining layers yields a multilayer network with $n=99$ countries and $L=30$ layers.

\begin{figure}[!tbp]
    \centering
    \includegraphics[width=0.6\linewidth]{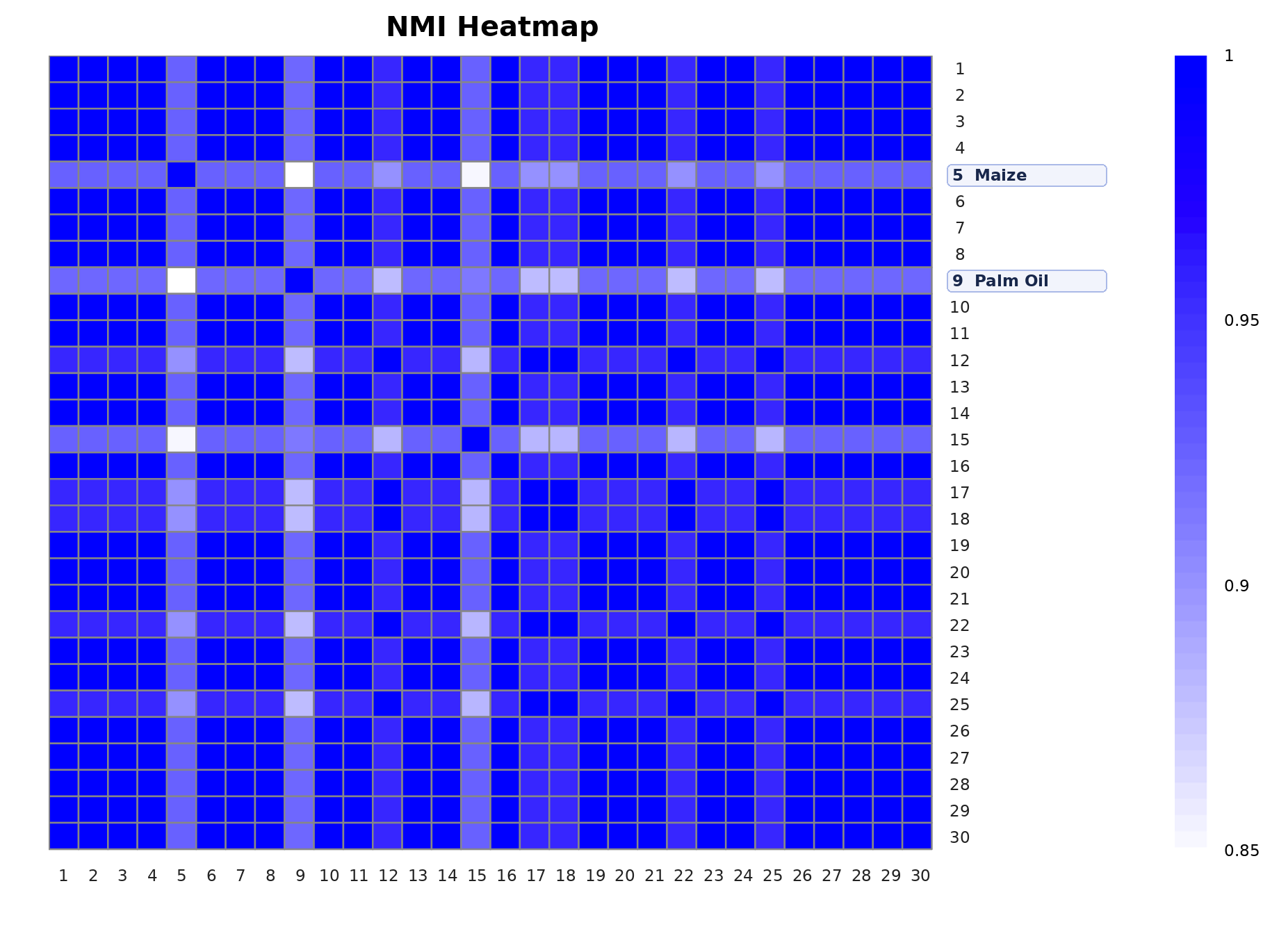}
    \caption{Pairwise NMI between the estimated layer-specific memberships for the 30 product layers.}
    \label{fig:foodnmi}
\end{figure}

We apply the proposed community-detection procedure in Algorithm~\ref{algo-lasc} with $(\hat K,K)=(6,3)$ to obtain the layer-specific memberships $\{\hat\pi^{(l)}\}_{l=1}^{30}$. For each layer pair $l, s$, we evaluate the similarity between their layer-specific membership estimates using the normalized mutual information (NMI). The corresponding $30 \times 30$ heatmap is presented in Figure~\ref{fig:foodnmi}. 
The estimated community structures are generally similar across product layers, with all pairwise NMI values above 0.85. This overall cross-layer agreement supports the use of information from other product networks in layer-specific community detection. Meanwhile, several layers exhibit noticeably lower NMI values with other layers, indicating product-specific trading structures. Examples of the latter include $l\in\{5,9,12,15,17,18,22,25\}$, consistent with the product heterogeneity in \citet{jing2021community}.

\begin{figure}[t]
    \centering
    \includegraphics[width=\linewidth]{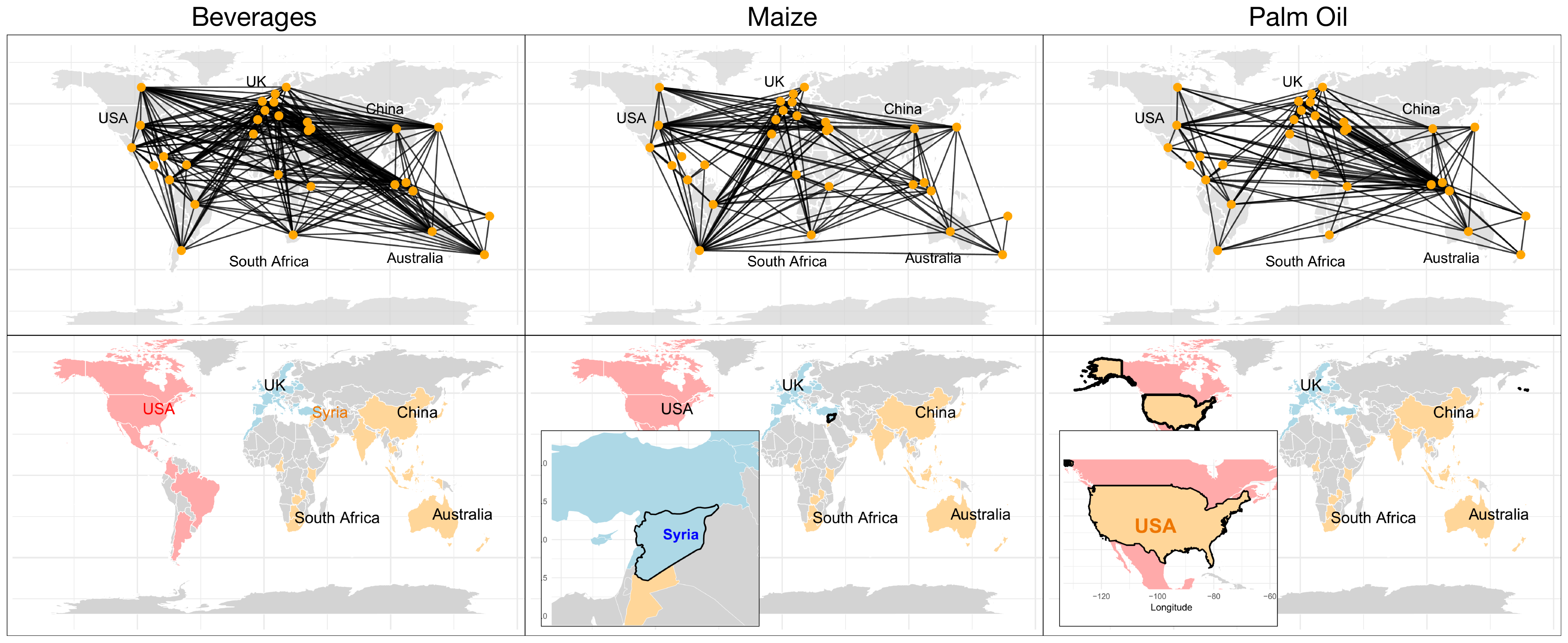}
    \caption{Connections (top row) and estimated community structures
(bottom row) for selected countries in three product networks.}
    \label{fig:foodre}
\end{figure}

We first examine the common structure shared by many product layers. A typical example is the beverages layer shown in the left panel of Figure~\ref{fig:foodre}. The three estimated communities largely correspond to the Americas, Asia \& Africa, and Europe. This geographic pattern is also reflected by the high-degree countries in each community, including Argentina, Brazil, and the United States in the Americas; China, India, and Japan in Asia \& Africa; and France, Germany, and the Netherlands in Europe. These results suggest that geography plays an important role in shaping trading activity.


By contrast, some layers emphasize product-specific roles. Figure~\ref{fig:foodre} shows the original network and community detection results for maize in the middle panel and palm oil in the right panel. For maize, Syria and Lebanon shift from their usual Asia \& Africa community to Europe, reflecting stronger trade links with Ukraine; for palm oil, the U.S. shifts toward Asia \& Africa through strong trade with Indonesia. These layer-specific labels characterize product-specific trading patterns and cross-layer heterogeneity. Overall, the food trading networks exhibit a common geographic structure together with meaningful product-specific heterogeneity, illustrating the value of layer-specific community detection.

\section*{Acknowledgments}
During the preparation of this work, the authors used
ChatGPT (OpenAI) to assist with improving the clarity,
language, and presentation of the manuscript.
After using this tool, the authors reviewed and edited
the content as necessary and take full responsibility
for the content of the publication.

\section*{Supplemental Materials}
\label{SM}
The supplementary materials contain all technical proofs and additional numerical analysis.



\bibliographystyle{plainnat}
\bibliography{reference}

\end{document}